\documentclass[10pt,journal,compsoc]{IEEEtran}

\usepackage[T1]{fontenc}
\usepackage[utf8]{luainputenc}
\usepackage[english]{babel}
\usepackage{array}
\usepackage{float}
\usepackage{epsfig}         
\usepackage{graphics}       
\usepackage{graphicx}
\usepackage{subfigure}
\usepackage{amsmath}
\usepackage{amssymb}
\usepackage{amstext}
\usepackage{amsfonts}
\usepackage{amssymb}
\usepackage{amsthm}
\usepackage{marvosym}
\usepackage{epstopdf}
\usepackage{verbatim}
\usepackage{cuted}
\usepackage{lipsum}
\usepackage{pdfpages}       
\usepackage{multicol}       
\usepackage{bm}
\usepackage{lineno}

\ifCLASSOPTIONcompsoc
  \usepackage[nocompress]{cite}
\else
  \usepackage{cite}
\fi

\ifCLASSINFOpdf
\else
\fi

\begin{document}
\title{TS-MPC for Autonomous Vehicles including a dynamic TS-MHE-UIO}

\author{Eugenio~Alcalá,~\IEEEmembership{}
        Vicenç~Puig~\IEEEmembership{}
        and~Joseba~Quevedo~\IEEEmembership{} \\
        \vspace*{5px}
\fontsize{8pt}{9pt}\selectfont  © 2018 IEEE. Personal use of this material is permitted. Permission from IEEE must be obtained for all other uses, in any current or future media, including reprinting/republishing this material for advertising or promotional purposes, creating new collective works, for resale or redistribution to servers or lists, or reuse of any copyrighted component of this work in other works.      
\IEEEcompsocitemizethanks{\IEEEcompsocthanksitem E. Alcalá, V. Puig and J. Quevedo are with the Department of Automatic Control, Polytechnic University of Catalonia, Barcelona, Spain.\protect\\
E-mail: eugenio.alcala@upc.edu} 
}

\IEEEtitleabstractindextext{%
\begin{abstract}


In this work, a novel approach is presented to solve the problem of tracking trajectories in autonomous vehicles.
This approach is based on the use of a cascade control where the external loop solves the position control using a novel Takagi Sugeno - Model Predictive Control (TS-MPC) approach and the internal loop is in charge of the dynamic control of the vehicle using a Takagi Sugeno - Linear Quadratic Regulator technique designed via Linear Matrix Inequalities (TS-LMI-LQR).
Both techniques use a TS representation of the kinematic and dynamic models of the vehicle.
In addition, a novel Takagi Sugeno estimator - Moving Horizon Estimator - Unknown Input Observer (TS-MHE-UIO) is presented. This method estimates the dynamic states of the vehicle optimally as well as the force of friction acting on the vehicle that is used to reduce the control efforts.
The innovative contribution of the TS-MPC and TS-MHE-UIO techniques is that using the TS model formulation of the vehicle allows us to solve the nonlinear problem as if it were linear, reducing computation times by 40-50 times.
To demonstrate the potential of the TS-MPC we propose a comparison between three methods of solving the kinematic control problem: using the non-linear MPC formulation (NL-MPC), using TS-MPC without updating the prediction model and using updated TS-MPC with the references of the planner.

\end{abstract}

}

\maketitle

\IEEEdisplaynontitleabstractindextext

\IEEEpeerreviewmaketitle

\ifCLASSOPTIONcompsoc
\IEEEraisesectionheading{\section{Introduction}\label{sec:introduction}}
\else
\section{Introduction}
\label{sec:introduction}
\fi

\IEEEPARstart{I}{n} the last recent years, we have experienced a great advance in the technological career towards autonomous driving. Today, we can see how research centers and large companies in the automotive sectors are accelerating and investing large amounts of money. In addition, if we add to this progress the advances in legislation and the increasing acceptance of the user, we converge on the fact that driving, as we know it today, has days countered.
The numerous advantages that the autonomous vehicle offers with respect to traditional vehicles are obvious. However, the most attractive is the great reduction of accidents on the roads, which will lead to a huge reduction in deaths on roads worldwide. \\

\noindent In order to achieve complete autonomous driving, a series of modules are needed working in a sequential and organized manner.
First, the vehicle sensing network (GPS, IMU, encoders, cameras, LIDAR, etc) collects all the information of the environment and the vehicle and is treated to extract measurements of interest (vehicle position and obstacles around, inertial measures, etc). Then, the trajectory planning module is responsible for generating the route using the real vehicle position and the desired one. This trajectory is composed of global positions, orientations and vehicle velocities.
Finally, it is the automatic control which, using this sequence of references and the position of the vehicle, generates the control actions (acceleration, steering and braking) for the actuators. \\

\noindent The automatic control is the last piece in the sequence of the autonomous vehicle and one of the most important tasks since it is in charge of its movement.
It is also the topic addressed in this paper.
The control problem can be defined by two general features: the type of control (lateral, longitudinal or integrated) and the type of model considered for its design (kinematic, linear dynamic, simplified non-linear dynamics or non-linear dynamics).
Articles \cite{jiang2018lateral} and \cite{yang2017algorithm} address the problem of lateral control using non-linear feedback control techniques.
Optimal-based techniques like LQR for lateral control problem is formulated in \cite{boyali2018tutorial}.
Regarding the longitudinal control, we can find LQR strategy in \cite{naeem2016autonomous,junaid2005lqr} and $H_2$ in \cite{naeem2016autonomous}.
However, these control strategies solve simplified versions of the real problem, i.e. the integrated control.
This work addresses both the longitudinal-lateral integrated control problem for autonomous vehicles. \\

\noindent Control strategies based on Linear Parameter Varying (LPV) and Takagi Sugeno (TS) models are techniques for solving non-linear problems using pseudo-linear models incorporating the non-linearities within the model parameters that depend on some scheduling variables.
The recent books, \cite{tanaka2004fuzzy}, \cite{gaspar10robust}, \cite{rotondo2017advances}, \cite{ostertag2011mono} and \cite{duan2013lmis}, presented the study of the modeling and design of LPV and TS under the formulation based on LMI.
Several design approaches can be used such as pole positioning, H$\infty$, H$_2$ and LQR.
These techniques have proven to be widely accepted in the field of robotics, for example \cite{rotondo2017advances} and \cite{blavzivc2017two}.

Another technique like Model Predictive Control (MPC) has proven to be one of the most interesting methods in this field in recent years. This strategy allows to find the optimal control action through the resolution of a constrained optimization problem in which a model of the system is evaluated in a future horizon.
Recent articles such as \cite{rawlings2017model}, \cite{corriou2018model} and \cite{mayne2014model} present the latest advances in MPC control outside the automotive field.
In the field of autonomous vehicles, we can find all kinds of formulations for the MPC.
From NLMPC applications in \cite{ercan2017adaptive}, where the lateral control problem is solved, to MPC lateral control using a linearized model of the vehicle in \cite{xu2017model}.
Working with nonlinear models give the best results. However, when working with systems with fast dynamics this technique may result non-viable since its excessive computational time.
That is why recent exploration of other ways opens the door to ideas such as mixing LPV-MPC or TS-MPC.
In \cite{cisneros2016efficient,besselmann2009autonomous} present the MPC strategy using LPV models.
In this work, the TS-MPC formulation is explored and presented in the field of autonomous vehicles. \\

\noindent The state measurement task will depend on the type of sensors installed on the vehicle and will be of vital importance for the application of certain control strategies.
The longitudinal and angular speeds are magnitudes easy to measure using low cost sensors and are common in many of today's vehicles. On the other hand, there are other variables that are more complicated or expensive to measure. That is why the estimation of the system states is a problem of great interest in the automotive field. \\
\noindent Regarding the external disturbances, one of the magnitudes that disturb with greater impact the dynamics of the vehicle is the friction force.
This one depends on the type of materials involved in the wheel-road contact. The rubber-asphalt contact is the most common and generates a magnitude of friction force that can be drastically altered if the vehicle suddenly crosses a wet or even frozen area.
For this reason, the estimation and subsequent compensation of this force by the control strategy is of great interest in the field of autonomous driving.
\noindent In recent years different approaches have been addressed in the area of disturbance estimation for autonomous vehicles such as \cite{yi1999estimation, svendenius2007tire, dakhlallah2008tire, alcala2018gain}.
One of these techniques is the Unknown Input Observer (UIO). This approach has been widely used for detection and isolation of faults \cite{amrane2017actuator} and presented in the autonomous vehicle field \cite{alcala2018gain}.
This type of observer allows to estimate the states of a system, as well as the disturbances or uncertainty not modeled in the system. \\

\noindent The contribution of this paper is twofold and focuses on the use of Takagi Sugeno polytopic models for the design of the control and observation stages.
First, the Model Predictive Control (MPC) technique is designed with a Takagi Sugeno (TS) kinematic model that allows the resolution of the optimization very quickly in comparison with current non-linear techniques. In addition, introducing the terminal set concept we are able to guarantee stability.
Second, the Moving Horizon Estimator (MHE) strategy is merged with the use of a dynamic vehicle model formulated as Takagi Sugeno (TS) as well as with the Unknown Input Observer (UIO) concept, thus allowing the estimation of states and disturbances through a very fast predictive optimization (TS-MHE-UIO). \\

\noindent The paper is structured as follows: Section 2 gives an overview of the work and describes the different types of modelling used for control and estimation purposes. In Section 3, the kinematic and dynamic control designs are developed. Section 4 shows the TS-MHE-UIO design.
Section 5 shows the simulation results and Section 6 presents the conclusions of the work.


\section{Overview of the Proposed Solution}

    The subject to be solved in this work consist on addressing the autonomous guidance problem of a city autonomous vehicle. To do so, two important tasks have to be carried out: the trajectory planning and the automatic control.

    On one hand, the planning of the trajectory to be followed by the vehicle has to fulfill certain specifications such as continuous and differentiable velocity profiles. Thus, this module is in charge of providing discrete and smooth references to the automatic control stage.

\begin{figure*}[htb!]
      \centering
      \includegraphics[width=160mm]{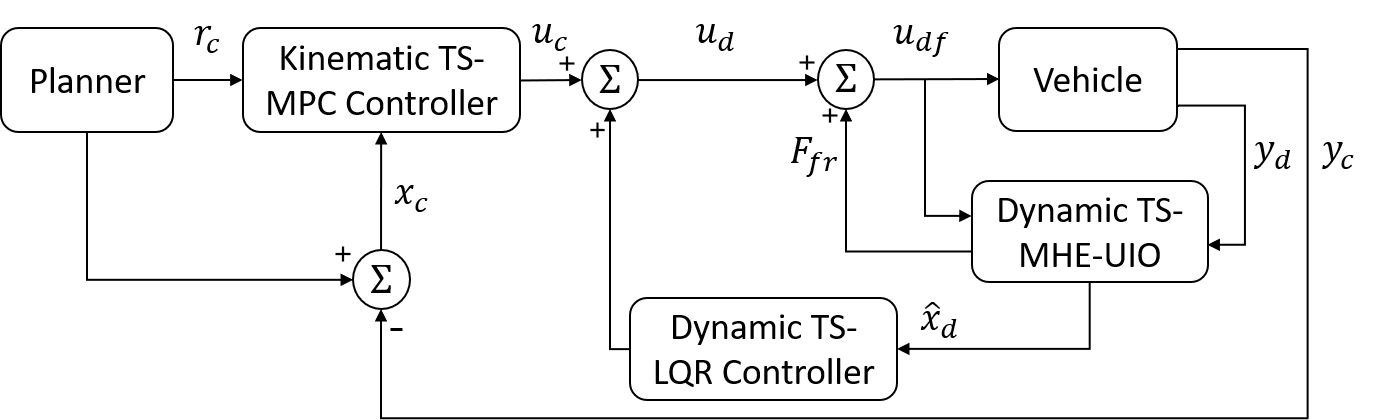}
      \caption{Autonomous guidance diagram with kinematic and dynamic control layers and a dynamic state estimator with friction force compensator.}
      \label{fig:control_scheme}
    \end{figure*}

    On the other hand, the automatic control is in charge of following the planned references, thus, moving the vehicle between two ground coordinates as well as generating smooth control actions for achieving a comfortable journey. In Fig. \ref{fig:control_scheme}, we show the planning-control-estimation diagram proposed in this work.
    Observe that two control levels have been designed, one for the position control and the other one to control the dynamic behaviour of the vehicle, i.e. velocities and slip angle.
    In addition, the lack of measurement of certain vehicle states as well as the ignorance of external disturbances can generate a problem when applying the designed control. Therefore, a dynamic estimator is introduced to solve this problem (see Fig. \ref{fig:control_scheme}).

    The level of difficulty of a vehicle guidance control problem comes often determined by two aspects: the type of control (lateral, longitudinal or mixed) and the complexity of the model to be controlled (kinematic, linear dynamic, non-linear simplified dynamic or non-linear dynamic).
    In this work we address the most complex configuration which is to solve the mixed non-linear dynamic problem. The following subsection covers the formulation of the different models used for solving the estimation and control problems.

\subsection{Takagi-Sugeno Control Oriented Models}
    When controlling mobile systems at low velocities the use of dynamic controls, and consequently dynamic models are not required. However, at higher velocities, i.e city cars, such a dynamic behaviour control becomes indispensable.

\begin{figure}[htbp]
      \centering
      \includegraphics[width=87mm]{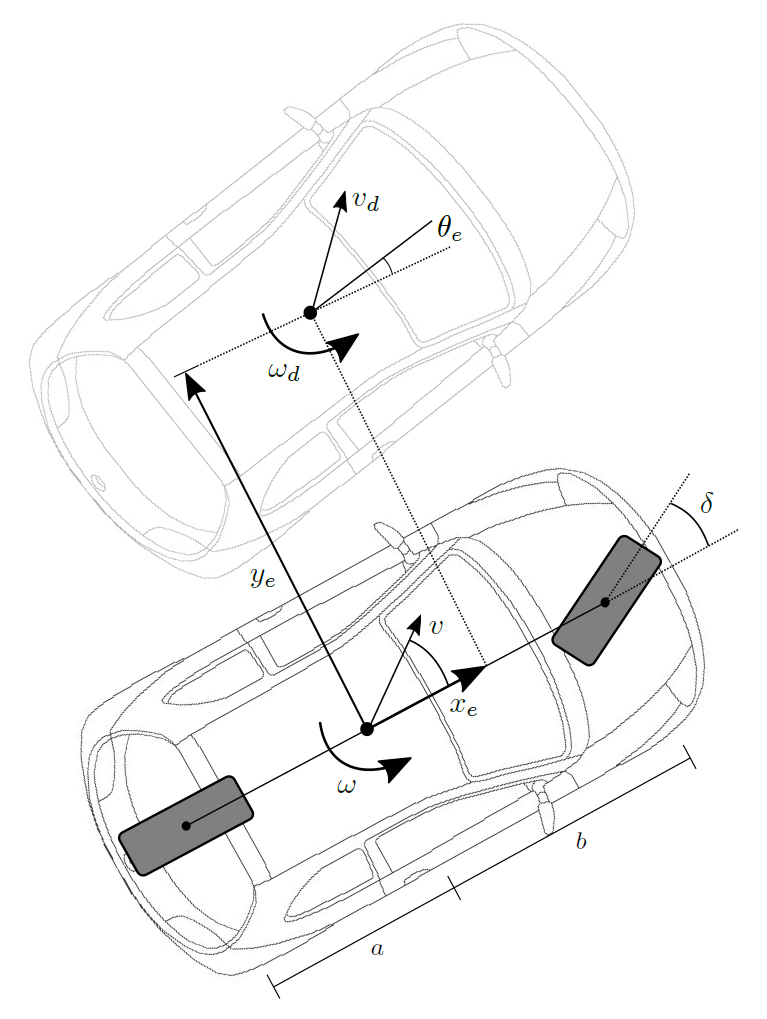}
      \caption{Two-wheels bicycle model used for estimation/control purposes. $\{ W \}$ frame represents the global inertial frame and $\{ B \}$ is the body frame located in the vehicle centre of mass.}
      \label{fig:vehicle_model}
    \end{figure}

    In this work, two model-based techniques cover the kinematic control and the dynamic estimation. For that reason, the use of mathematical kinematic and dynamic vehicle models are necessary.
    The kinematic model is based on the mass-point assumption while for the dynamic one, the two-wheels bicycle model has been considered as the one presented in  Fig.\ref{fig:vehicle_model}.

    \noindent In the following subsections, we present the TS formulation of the kinematic and dynamic models. The full development of these models is presented in Sections 2 and 3 in \cite{alcala2018gain}.

\subsubsection{Kinematic TS model}

    Denoting the state, control and reference vectors, respectively, as

    \begin{subequations}
    \label{eq:kinematic_TS_model}
        \begin{equation}
            \label{eq:kinematic_state_space_vectors}
            x_c =
            \left[\begin{array}{c}
                x_e \\
                y_e  \\
                \theta_e  \\
            \end{array}\right] , \
            u_c =
        	\left[\begin{array}{c}
                v \\
                \omega \\
            \end{array}\right] , \
            r_c =
                \left[\begin{array}{c}
                     v_d \cos \theta_e \\
                     \omega_d \\
            \end{array}\right]
        \end{equation}

        \noindent and defining the vector of scheduling variables as $\rho(k) := \left[ \omega(k), v_d(k), \theta_e(k) \right]$ which are bounded in $\omega \in [-1.42, 1.42]$, $v_{d} \in [0.1, 20]$ and $\theta_e \in [-0.05, 0.05]$, then, the non-linear kinematic model (see Chapter 1 of \cite{alcala2018autonomous}) is transformed into the Takagi-Sugeno representation by using the sector nonlinearity approach
        \begin{equation}
                x_c(k+1) = A_c(\rho(k)) x_c(k) + B_c u_c(k) - B_c r_c(k) \\
        \end{equation}

        \noindent where
        \begin{equation}
            A_c \left( \rho(k) \right) =
        	\left[\begin{array}{ccc}
                1 & \omega T_c & 0 \\
                -\omega T_c & 1 & v_d \frac{\sin \theta_e}{\theta_e} T_c \\
                0 & 0 & 1 \\
            \end{array}\right]
        \end{equation}
        \begin{equation}
            B_c =
            \left[\begin{array}{cc}
                -1 & 0 \\
                 0 & 0 \\
                 0 & -1 \\
            \end{array}\right] T_c .
        \end{equation}
    \end{subequations}
    with $T_c$ being the kinematic sample time.

    \noindent From this formulation, a polytopic representation for the control design is obtained as

    \begin{equation}
    \label{eq:polytopic_kinematic_system}
        x_c(k+1) = \sum_{i=1}^{2^{r_{c}}} \mu_i (\rho(k)) A_{c_i} x_c(k) + B_c u_c(k) - B_c r_c(k)
    \end{equation}
    being $r_c$ the number of scheduling variables and $A_i$ each one of the polytopic vertex systems obtained as a combination of the extreme values of the scheduling variables.

    \noindent The expression $\mu_i (\rho(k))$ is known as the membership function and is given by

    \begin{subequations}
    \label{eq:weighting_function}
        \begin{equation}
            \mu_{i}(\rho(k)) = \prod_{j=1}^{r_c} \xi_{ij}(\eta_{0}^{j}, \eta_{1}^{j}) \ , \ \ \ \ i= \{ 1,...,2^{r_c} \}
        \end{equation}

        \begin{equation}
        \begin{aligned}
            & \eta_{0}^{j} = \frac{ \overline{\rho_{j}} - \rho_{j}(k)  }{\overline{\rho_j} - \underline{\rho_j}} \\
            & \eta_{1}^{j} = 1 - \eta_{0}^{j}   \ , \ \ \ \ \ \ \ \ \ \ j = \{1,...,r_c\}
        \end{aligned}
        \end{equation}

        \noindent where $\xi_{ij}(\eta_{0}^{j}, \eta_{1}^{j})$ corresponds to any of the weighting function that depend on each rule $i$.
    \end{subequations}

\subsubsection{Dynamic TS model}

    The dynamic TS model considered in this work is a transformation of the non-linear one presented in \cite{alcala2018gain}.

    \noindent Then, denoting the state and control vectors, respectively, as

    \begin{subequations}
    \label{eq:dynamic_TS_model}
        \begin{equation}
            \label{eq:dynamic_state_space_vectors}
            x_d =
            \left[\begin{array}{c}
                v \\
                \alpha  \\
                \omega  \\
            \end{array}\right] , \
            u_d =
        	\left[\begin{array}{c}
                F_{xR}\\
                \delta \\
            \end{array}\right]
        \end{equation}

        \noindent and considering an unknown friction force disturbance $F_{fr}$ as a variation of the nominal friction force ($\mu_o M g$) ,the Takagi-Sugeno model can be expressed as
        \begin{equation}
                x_d(k+1) = A_d (\vartheta(k)) x_d(k) + B_d (\vartheta(k)) u_d(k) + E_d F_{fr}(k) \\
        \end{equation}

        \noindent where $\vartheta(k) := [\delta(k), v(k), \alpha(k)]$ with $\delta \in [-1.42, 1.42]$, $v \in [0.1, 20]$ and $\alpha \in [-0.1, 0.1]$; and

        \begin{equation}
            A_d (\vartheta(k)) =
        	\left[\begin{array}{ccc}
                1+A_{11}T_d & A_{12}T_d & A_{13}T_d \\
                0 & 1+A_{22}T_d & A_{23}T_d \\
                0 & A_{32}T_d & 1+A_{33}T_d  \\
            \end{array}\right]
        \end{equation}
        \begin{equation}
            A_{11} = -\frac{\frac{1}{2} C_d \rho A_r v^2 + \mu_o M g}{Mv} \\
        \end{equation}
        \begin{equation}
            A_{12} = \frac{C_{x} ( \sin(\delta)\cos(\alpha) - \sin(\alpha)\cos(\delta) -\sin(\alpha) ) }{M} \\
        \end{equation}
        \begin{equation}
            A_{13} = \frac{C_{x} (a ( \sin(\delta)\cos(\alpha)-\sin(\alpha)\cos(\delta) ) + b \sin(\alpha) ) }{M v} \\
        \end{equation}
        \begin{equation}
            A_{22} = \frac{-C_{x} (\cos(\alpha)\cos(\delta)+\sin(\alpha)\sin(\delta)+\cos(\alpha)) }{M v} \\
        \end{equation}
        \begin{equation}
            A_{23} = \frac{ -C_{x} a (\cos(\delta)\cos(\alpha)+\sin(\alpha)\sin(\delta)) + C_{x} b \cos(\alpha) }{M v^2}-1 \\
        \end{equation}
        \begin{equation}
            A_{32} = \frac{C_{x} ( b - a \cos(\delta) )}{I} \ , \
            A_{33} = \frac{-C_{x} (b^2 + a^2 \cos(\delta) ) }{I v} \\
        \end{equation}

        \begin{equation}
            \label{eq:matrices_state_space_B_D}
            B_d (\vartheta(k)) =
            \left[\begin{array}{cc}
                B_{11} & B_{12} \\
                B_{21} & B_{22} \\
                0 & B_{32} \\
            \end{array}\right] T_d
        \end{equation}

        \begin{equation}
            B_{11} = \frac{\cos(\alpha)}{M} \ , \
            B_{12} = \frac{C_{x} ( -\sin(\delta)\cos(\alpha)+sin(\alpha)\cos(\delta))}{M} \\
        \end{equation}
            \begin{equation}
            B_{21} = \frac{-\sin(\alpha)}{M v} \ , \
            B_{22} = \frac{C_{x} (\cos(\alpha)\cos(\delta)+\sin(\alpha)\sin(\delta)) }{M v} \\
        \end{equation}
            \begin{equation}
            B_{32} = \frac{C_{x} a \cos(\delta)}{I}
        \end{equation}
        \begin{equation}
            \label{eq:matrices_state_space_C_D}
            E_d =
            \left[\begin{array}{c}
                 \frac{-T_d}{M} \\
                 0 \\
                 0 \\
            \end{array}\right]
        \end{equation}
    \end{subequations}
    with $T_d$ being the sample time using in the dynamic control.

    \noindent As in the case of kinematic model, we look for a polytopic dynamic formulation like the following
    \begin{equation}
    \label{eq:polytopic_dynamic_system}
        x_d(k+1) = \sum_{i=1}^{2^{r_{d}}} \mu_i \big( \vartheta(k)\big) \big(A_{d_i} x_d(k) + B_{d_i} u_d(k) \big) + E_d F_{fr}(k)
    \end{equation}
    being $r_d$ the number of dynamic scheduling variables and $A_{d_i}$ and $B_{d_i}$ represent each one of the polytopic vertex dynamic systems obtained as a combination of the extreme values of the dynamic scheduling variables. The membership function is the same than the one presented in \eqref{eq:polytopic_kinematic_system} but using the dynamic scheduling vector $\vartheta(k)$.

\section{Control Design}
In this section, we present the control scheme proposed for this work as well as its design.
The control strategy of the vehicle has been divided into two nested layers, see Fig \ref{fig:control_scheme}.
The outermost layer controls the vehicle's kinematics, i.e. its position and orientation, and works at a frequency of 20 Hz.
On the other hand, the internal loop controls the dynamic behavior of the vehicle, i.e. its speeds, at a frequency of 200 Hz.
Next, both control loops are described separately.

\subsection{Kinematic TS-MPC Design}
    At this point, we present the formulation of the Takagi Sugeno Model Predictive Control strategy, which focuses on solving position and orientation control of the vehicle.

    \noindent This strategy is based on the resolution of a linear quadratic optimization problem by using the non-linear kinematic error model in its TS polytopic representation.
    However, there exist the problem associated with the lack of knowledge of the matrix of scheduling variables through the entire prediction horizon.
    In \cite{cisneros2016efficient}, the use of the optimized state sequence which is obtained after each optimization is proposed.

    In this work, the scheduling variables are states of the system whose desired values are known since the trajectory planner generates them.
    That is why we propose the use of such references as known scheduling variables for the entire optimization horizon being then the scheduling sequence $\Gamma := [\rho(k), ..., \rho(k+N)]$. In this way, we can calculate the evolution of the model more accurately and in anticipation.

    \noindent In addition, since the basic MPC formulation cannot guarantee the overall stability of the system, we propose the addition of a terminal constraint and a terminal cost to the optimization problem.

    \noindent To formulate the problem, the polytopic TS system presented in \eqref{eq:polytopic_kinematic_system} has been considered.
    In order to avoid a difficult reading the sub-index $c$ is omitted in the rest of the subsection.
    \noindent Then, the focus is on a model predictive control scheme where the cost function is defined as

    \begin{equation}
        \mathrm{J_k}= \sum_{i=0}^{N-1} \big( x_{k+i}^T Q x_{k+i} + \Delta u_{k+i} R \Delta u_{k+i} \big)
        + x_{k+N}^T P x_{k+N}
    \end{equation}

    \noindent where $Q = Q^T \ge 0$, $R = R^T > 0$, $P = P^T > 0$ and the $N+1$ term represents the terminal cost.

    \noindent At each time $k$ the values of $x_k$ and $u_{k-1}$ are known and the following optimization problem can be solved

    \begin{equation}
    \label{eq:MPC}
    \begin{aligned}
    & \underset{\Delta U_k}{\text{minimize}}
    && \mathrm{J_k}(\Delta U_k, X_k)   \\
    & \text{subject to}\\
    &&& x_{k+i+1} = \sum_{j=1}^{2^{r_c}} \mu_j (\rho_{k+i}) A_j x_{k+i} + B u_{k+i} - B r_{k+i} \\
    &&& u_{k+i} = u_{k+i-1} + \Delta u_{k+i} \ \, \ \ \ \ i=0,...,N-1\\
    &&& \Delta U_k \in \Delta \Pi \\
    &&& U_k \in \Pi  \\
    &&& x_{k+N} \in \chi \\
    \end{aligned}
    \end{equation}

    \noindent where
    \begin{equation}
        \Delta U_k = \left[\begin{array}{c}
                    \Delta u_k     \\
                    \Delta u_{k+1} \\
                    \vdots         \\
                    \Delta u_{k+N-1} \\
        \end{array}\right] \in \mathbb{R}^{m}  \ , \
        U_k = \left[\begin{array}{c}
                    u_k     \\
                    u_{k+1} \\
                    \vdots  \\
                    u_{k+N-1} \\
        \end{array}\right] \in \mathbb{R}^{m}
    \end{equation}


    \noindent being $m$ the number of inputs of the kinematic system. $\Pi$ and $\Delta \Pi$ are the constraint sets for the inputs and their derivatives, respectively.

    \noindent The set $\chi$ represents the terminal state domain. Then, by introducing this constraint in the optimization problem, we force the states to converge into a stable region and then ensure the MPC stability.
    The computation of this terminal set is carried out by solving two LMI-based problems.
    First a controller $K_i = W_i Y^{-1}$ for each polytopic system ($A_i$) is found by solving the following LQR-LMI

    \begin{equation}
    \label{eq:Feasibility-LMI}
    \begin{aligned}
        & \left[\begin{array}{cccc}
          	    Y     &  (A_iY+BW)^T &  Y     &  W_i^T \\
         	    A_iY+BW &    Y       &  0     &  0  \\
         	    Y     &    0       & Q_{TS}^{-1} &  0  \\
         	    W_i     &    0       &  0     &  R_{TS}^{-1}
                    \end{array}\right] \ge 0 \\
        & i = 1,...,2^{r_c}
    \end{aligned}
    \end{equation}
    with $Y=Y^T>0$, $Q_{TS} = Q_{TS}^T \ge 0$, $R_{TS} = R_{TS}^T > 0$.

    \noindent This LQR design is a particular formulation for the one presented in Theorem 25 of \cite{tanaka2004fuzzy}. The constant nature of kinematic input matrix $B_c$ in \eqref{eq:kinematic_TS_model} allows the use of this simplified LMI version.

    \noindent The second problem consists on finding the largest terminal region $\chi = \{ x|x^T S x \le 1 \}$, with $S=Z^{-1}$, by solving the following constrained optimization problem

    \begin{equation}
    \label{eq:terminal_set_computation}
    \begin{aligned}
    & \underset{Z}{\text{maximize}}
    & & \mathrm{J_k}(Z)   \\
    & \text{subject to} \\
    && \left[\begin{array}{cc}
          	    -Z     &  Z(A_i+BK_i)^T \\
         	    (A_i+BK_i)Z &    -Z
            \end{array}\right] < 0 \\
    && K_i Z K_i^T - \overline{u}^2 < 0  \ \ \ \ \  i = 1,...,2^{r_c}
    \end{aligned}
    \end{equation}

\subsection{Dynamic TS-LQR Design}
\label{sec:TS-LQR}


    \noindent It is important to emphasize that the fact that the input matrix $B_d$ in \eqref{eq:dynamic_TS_model} is parameter varying makes the LMI formulation presented in \eqref{eq:Feasibility-LMI} more complex.
    At this point, there are two ways to deal with this problem.
    The first one is using the LMI formulation extended to the case of B variant as presented in the Theorem 25 of \cite{tanaka2004fuzzy}.
    The second is to expand the model using a filter as proposed in \cite{Apkarian1994}. With this last method, we obtain a system with more states but with a constant input matrix, thus being able to use the formulation in \eqref{eq:Feasibility-LMI}.

    \noindent Thereupon, to design the controller gain, we use the polytopic system \eqref{eq:polytopic_dynamic_system} being matrices $A_d$, $B_d$ and $E_d$ the augmented matrices presented in (17) in \cite{alcala2018gain}.
    Then, we solve the optimal LMI problem \eqref{eq:Feasibility-LMI} for computing the polytope vertex controllers $K_i$.

    \noindent Hence, the dynamic controller gain at each control iteration is given by
    \begin{equation}
        \label{eq:interpolation_K}
        K (\vartheta (k)) = \sum_{i=1}^{2^{r_d}} \mu_i (\vartheta (k)) K_i
    \end{equation}

    \noindent where $\mu_i(\vartheta (k))$ represents the weighting function presented in \eqref{eq:weighting_function} by using the dynamic scheduling vector.

    \noindent The offline computation of polytope controllers allows this control strategy to execute at the desired frequency of 200 Hz.

\section{TS-MHE-UIO Design}
    On one hand, the aim of the Moving Horizon Estimator is to predict the dynamic states for the next iteration by means of running an optimization and using a set of past allowed measurements.
    On the other hand, the Unknown Input Observer deals with the estimation of external disturbances.
    One of the most relevant disturbances in road vehicles is the continuous change of road materials.
    That is why the coefficient of friction varies producing a remarkable alteration in the total computation of acting forces, drastically affecting the behavior of the vehicle.

    In this section, we present a novel combining both the MHE and the UIO, to converge to an optimal state estimator able to predict disturbances.
    In addition, using a TS model formulation for computing the evolution during the established horizon allows the algorithm to run faster than non-linear model-based MHE.

    \noindent To avoid a difficult reading the sub-index $d$ is omitted in system vectors but not in systems matrices.

    \subsection{UIO}

    The UIO goal is to estimate the main disturbances acting over the vehicle. Such a procedure is based on calculating the difference between the observation model and the real system \cite{Keller1999}. In this work, we have considered as the disturbance the friction force
        \begin{equation}
    \label{eq:disturb_estimation}
        F_{fr}(k) = \Theta \Big(y(k) - C_d \big( \sum_{i=1}^{2^r} \mu_i (\vartheta_k) A_{d_i} \hat{x} + B_d u \big) \Big) ,
    \end{equation}

    \noindent with

    \begin{equation}
        \label{eq:observability_C_matrix}
        C_d =
            \left[\begin{array}{ccc}
                 1 & 0 & 0  \\
                 0 & 0 & 1
            \end{array}\right] \ , \
        \hat{x} =
        \left[\begin{array}{c}
            \hat{v} \\
            \hat{\alpha}  \\
            \hat{\omega}  \\
        \end{array}\right] \ , \
        \Theta = (C_d E_d)^{+} \ ,
    \end{equation}

    \noindent where $A_{d_i}$, $B_d$ and $E_d$ are the dynamic system matrices in \eqref{eq:dynamic_TS_model} and $\mu_i(\vartheta_k)$ represents the weighting function presented in \eqref{eq:weighting_function} by using the dynamic scheduling vector $\vartheta_k$ defined in \eqref{eq:weighting_function}.

    \noindent Then, at every control iteration and once the state estimation has been solved, $F_{fr}$ is computed.

    \subsection{TS-MHE Design}

    In order to design the MHE-UIO the polytopic TS system \eqref{eq:polytopic_dynamic_system} is used.
    Taking into account the availability of the sensors available on current vehicles, we can consider the slip angle state as an unmeasured variable which motivates the use of a state estimator.
    The MHE problem is based on minimizing the following cost function
    \begin{equation}
        \mathrm{J_k} = (\hat{x}_{k-N}-x_o)^T P (\hat{x}_{k-N}-x_o)
    \end{equation}
    \begin{equation*}
        + \sum_{i=-N}^{0} \big( w_{k+i}^T Q w_{k+i} + s_{k+i}^T R s_{k+i} \big).
    \end{equation*}

    \noindent Therefore, at each instant of time $k$, knowing the vectors

    \begin{equation}
        U_k = \left[\begin{array}{c}
                    u_{k-N}     \\
                    u_{k-N+1} \\
                    \vdots         \\
                    u_{k} \\
        \end{array}\right] \in \mathbb{R}^{m}  \ , \
        Y_k = \left[\begin{array}{c}
                    y_{k-N}     \\
                    y_{k-N+1} \\
                    \vdots         \\
                    y_{k} \\
        \end{array}\right] \in \mathbb{R}^{m}
    \end{equation}

    \noindent and the initial state $x_o$, the constrained optimization problem
    \begin{equation}
    \label{eq:MHE}
    \begin{aligned}
    & \underset{\hat{X}_k}{\text{minimize}}
    & & \mathrm{J_k}(\hat{X}_k)   \\
    & \text{subject to}
    & & \hat{x}_{k+i+1} = \sum_{j=1}^{2^{r_d}} \mu_j (\vartheta_{k+i}) \big( A_{o_{j}} \hat{x}_{k+i} + B_{o_{j}} u_{k+i} \big) \\
    &&& + w_{k+i} + E_d \Theta y_{k+i} \,\,\,\, i=-N,...,0\\\\
    &&& y_{k+i} = C_d \hat{x}_{k+i} + s_{k+i} \,\,\,\, i=-N,...,0\\
    &&& \hat{X}_k \in X_d \\
    \end{aligned}
    \end{equation}

    \noindent is solved online for
    \begin{equation}
            \hat{X}_k = \left[\begin{array}{c}
                    \hat{x}_{k-N+1}     \\
                    \hat{x}_{k-N+2} \\
                    \vdots  \\
                    \hat{x}_{k+1} \\
        \end{array}\right]    \in \mathbb{R}^{n_x} \ ,
    \end{equation}

    \noindent where $X_d$ is the constraint region for the dynamic states, $n_x$ is the number of dynamic states, $Q = Q^T \ge 0$, $R = R^T > 0$, $P = P^T > 0$ and

    \begin{equation*}
            A_{o_{j}} = (I - E_d \Theta C_d) A_{d_j}
    \end{equation*}
    \begin{equation*}
            B_{o_j} = (I - E_d \Theta C_d) B_{d_j} .
    \end{equation*}

\noindent are the unknown input matrices \cite{Keller1999}.

\section{Simulation Result}
In this section, we validate the performance of the presented control/observer scheme through simulation in MATLAB.
To show the promising results of the TS-MPC, we perform a comparison against the non-linear MPC approach (NL in resulting figures). Moreover, we compare two different methods for evaluating the MPC model in the prediction stage. One based on freezing the system along the prediction horizon and another based on using the desired references that the planner provides (FZN and REF in resulting figures).
Then, the optimal control problem \eqref{eq:MPC} is solved at every 100 ms using the solver GUROBI \cite{optimization2014inc} through YALMIP \cite{lofberg2004yalmip} framework. This solves the position control problem in a outer loop. In the inner loop the dynamic state feedback control problem (Section \ref{sec:TS-LQR}) is solved at every 10 ms to control the velocities of the vehicle.
In addition, the states used by this inner control law are provided by the optimal observer \eqref{eq:MHE}.
The vehicle model, TS-MPC, TS-MHE-UIO and dynamic TS-LQR parameters are listed in Tables \ref{table:vehicle_parameters}, \ref{table:MPC_parameters}, \ref{table:MHE_parameters} and \ref{table:LQR_parameters}, respectively.

    \begin{table}[htbp]
    \caption{Dynamic model parameters.}
    \label{table:vehicle_parameters}
    \begin{tabular}{ l|l||l|l }
    \hline
    Parameter & Value & Parameter & Value \\
    \hline
    \hline
    $a$     & 0.758 $m$ & $A_r$   & 1.91 $m^2$  \\
    $b$     & 1.036 $m$ & $\rho$  & 1.184 $\frac{kg}{m^3}$  \\
    $M$     & 683 $kg$  & $C_d$   & 0.36  \\
    $I$     & 560.94 $kg$ $m^2$   & $\mu_o$ & 0.5  \\
    $C_x$   & 25000 $\frac{N}{rad}$ \\
    \hline
    \end{tabular}
    \end{table}

    \begin{table}[htbp]
    \caption{TS-MPC design parameters.}
    \label{table:MPC_parameters}
    \begin{tabular}{ l|l||l|l }
    \hline
    Parameter & Value & Parameter & Value \\
    \hline
    \hline
    $Q$     & diag(1.133 0.067 13.333)   & $\overline{u}$ & [18 1.4]  \\
    $R$     & diag(0.000005 5.5)    & $\underline{u}$    & [0 -1.4]  \\
    $T_c$   & 0.1 $s$    & $\overline{\Delta u}$  & [5 0-3]  \\
    $N$     & 40    & $\underline{\Delta u}$ & [-5 -0.3]  \\
    $R_{TS}$ & diag(1 3)    & $Q_{TS}$               & diag(1 1.5 3)   \\
    \hline
    \end{tabular}
    \end{table}

    \begin{table}[htbp]
    \caption{TS-MHE-UIO design parameters.}
    \label{table:MHE_parameters}
    \begin{tabular}{ l|l||l|l }
    \hline
    Parameter & Value & Parameter & Value \\
    \hline
    \hline
    $Q$     & diag(10 10 2)     & $N$     & 30 \\
    $R$     & diag(0.033 0.033) & $T_d$   & 0.01 $s$ \\
    $P$     & diag(2 2 2)  & $\overline{\hat{x}}$ & [ 18 0.1 1.4] \\
    $\underline{\hat{x}}$ & [ 0 -0.1 -1.4] \\
    \hline
    \end{tabular}
    \end{table}

    \begin{table}[htbp]
    \caption{TS-LQR design parameters.}
    \label{table:LQR_parameters}
    \begin{tabular}{ l|l }
    \hline
    Parameter & Value \\
    \hline
    \hline
    $Q$     & diag(2500 0.1 0.1 0.1 100000 90000)   \\
    $R$     & diag(0.001 9)                         \\
    \hline
    \end{tabular}
    \end{table}

\noindent Solving the problem \eqref{eq:terminal_set_computation} for computing the largest terminal set, we obtain matrix $S$ as
\begin{equation}
     S = \left[\begin{array}{ccc}
                 0.465 & 0 & 0  \\
                 0 & 23.813 & 76.596 \\
                 0 & 76.596 & 257.251
         \end{array}\right] .
\end{equation}

\begin{figure}[htbp]
      \centering
      \includegraphics[width=95mm]{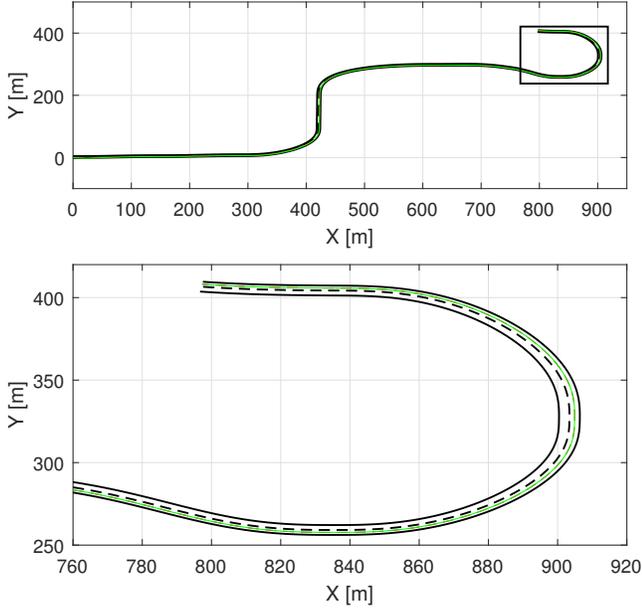}
      \caption{Complete simulation circuit on top. Zoom of a part of the circuit below. }
      \label{fig:circuit}
    \end{figure}

\begin{figure}[htbp]
      \centering
      \includegraphics[width=95mm]{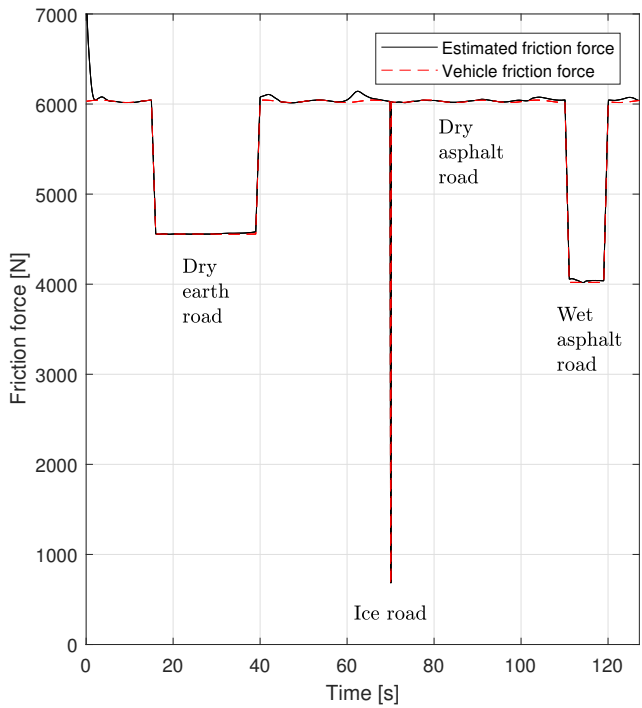}
      \caption{Friction force profile applied during the simulation test and the estimated friction force by the TS-MHE-UIO.}
      \label{fig:friction_force_profile}
    \end{figure}

\noindent The comparison is made in the circuit presented in Fig. \ref{fig:circuit} under the influence of disturbance in the friction coefficient. It is intended to show a real case in which the vehicle passes through dry and wet asphalt surfaces, dry earth surface and even frozen surface (see Fig. \ref{fig:friction_force_profile}).

\begin{figure}[htbp]
      \centering
      \includegraphics[width=95mm]{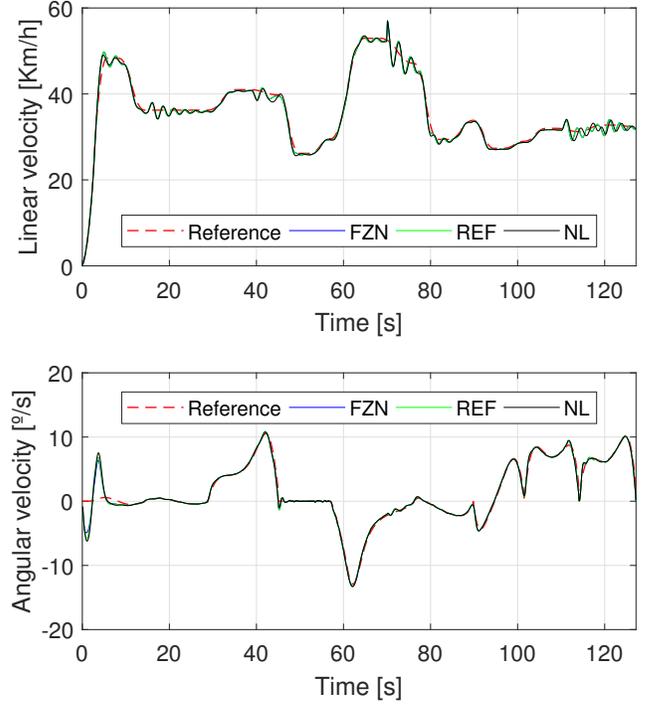}
      \caption{Reference and response velocities for each one of the compared methods.}
      \label{fig:velocity_results}
    \end{figure}

\begin{figure}[htbp]
      \centering
      \includegraphics[height=100mm]{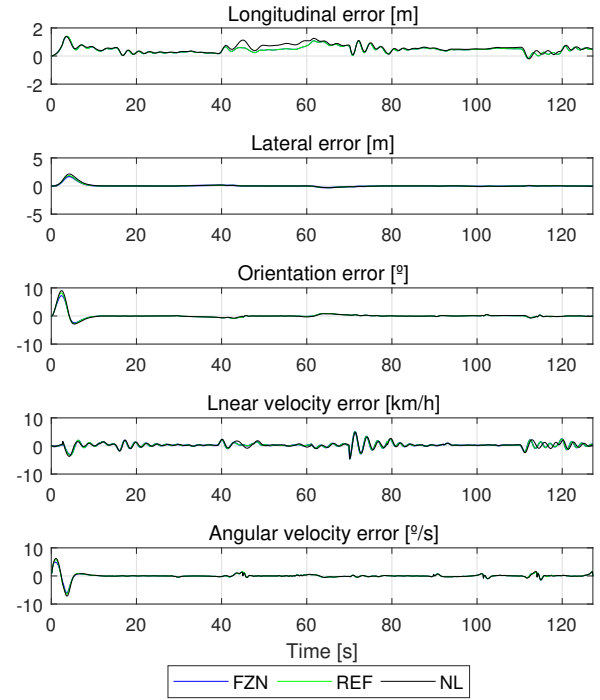}
      \caption{Kinematic and dynamic vehicle errors for each compared kinematic control strategy.}
      \label{fig:errors_results}
    \end{figure}

\begin{figure}[htbp]
      \centering
      \includegraphics[width=95mm]{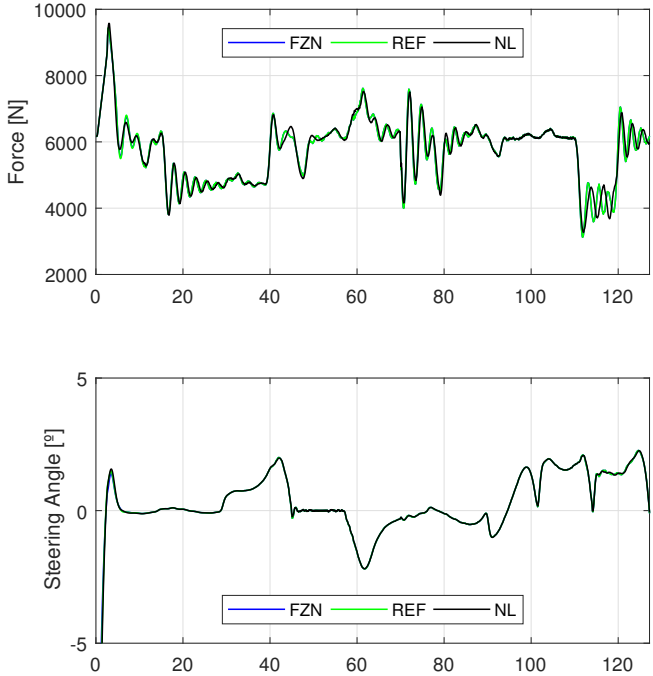}
      \caption{Resulting control actions for the three compared methods.}
      \label{fig:control_actions_results}
    \end{figure}

\noindent Below, in Fig. \ref{fig:velocity_results}, we show both, the linear and angular speed profiles provided by the trajectory planning and the respective vehicle responses.
\noindent In Fig. \ref{fig:errors_results}, we illustrate the complete set of errors, i.e. position, orientation and velocities errors. It can be seen that there is hardly any difference in the results between the techniques compared.
The respective control actions applied to the simulation vehicle are shown in Fig \ref{fig:control_actions_results}
and the friction force affecting the vehicle as well as the estimation of this by the TS-MHE-UIO are depicted in Fig \ref{fig:friction_force_profile}.

\begin{figure}[htbp]
      \centering
      \includegraphics[width=95mm]{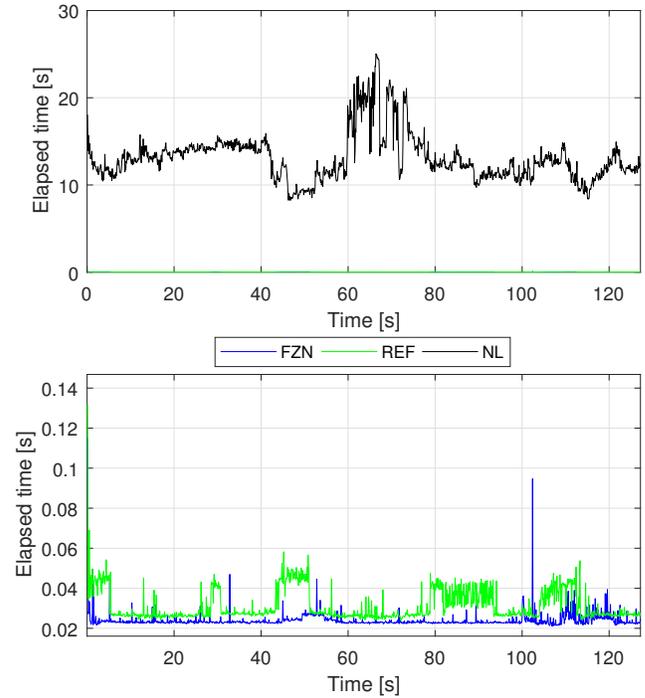}
      \caption{Computational time when solving the kinematic MPC strategy with the different methods. }
      \label{fig:elapsed_time_results}
    \end{figure}

\noindent An important aspect of control strategies based on optimization is the computational time spent at each optimization procedure. In Fig \ref{fig:elapsed_time_results} we show the used time at each kinematic MPC optimization for each one of the three compared methods.
Finally, a quantitative comparison is made using the root mean squared error (RMSE) as performance measurement. It is shown in Table \ref{table:RMSE}.
    \begin{table}[htbp]
    \caption{Comparison using a quadratic measure.}
    \label{table:RMSE}
    \begin{tabular}{ l|l|l|l|l|l }
    \hline
    Approach & $RMSE_x$ & $RMSE_y$ & $RMSE_{\theta}$ & $RMSE_v$ & $RMSE_w$ \\
    \hline
    \hline
    $FZN$  & 0.517 & 0.259 & 0.017 & 0.258 & 0.014  \\
    $REF$  & 0.501 & 0.238 & 0.016 & 0.251 & 0.013  \\
    $NL$   & 0.528 & 0.225 & 0.015 & 0.268 & 0.012  \\
    \hline
    \end{tabular}
    \end{table}




\section{Conclusion}

In this work, a cascade control scheme (kinematic and dynamic) was presented to solve the problem of integrated control (lateral and longitudinal) of autonomous vehicles.

\noindent The novel kinematic control was designed using the Model Predictive Control technique with the prediction model expresed in the Takagi Sugeno formulation (TS-MPC). On the other hand, the dynamic control was approached using the Linear Quadratic Regulator strategy, with a Takagi Sugeno modeling and using a LMI formulation of the problem (TS-LMI-LQR).

\noindent A comparison was made between three methods of solving the control problem: using the non-linear MPC formulation (NL-MPC), using TS-MPC without updating the prediction model and using updated TS-MPC with the planner references.
It was demonstrated that the TS-MPC technique works very well compared to the non-linear control problem but in a much faster way (between 40 and 50) times faster.

\noindent In addition, a Takagi Sugeno - Moving Horizon Estimator - Unknown Input Observer (TS-MHE-UIO) was introduced with the aim of estimating dynamic states and disturbances acting on the vehicle, such as the friction force. The estimation of the friction force is used to compensate the disturbance and allow lower control efforts.

As future research, it is planned to apply the proposed strategy in a real vehicle.



\ifCLASSOPTIONcompsoc
  \section*{Acknowledgments}
\else
  \section*{Acknowledgment}
\fi

This work has been funded by the Spanish Ministry of Economy and Competitiveness (MINECO) and FEDER through the projects SCAV (ref. DPI2017-88403-R) and HARCRICS (ref. DPI2014-58104-R). The author is supported by a FI AGAUR grant (ref 2017 FI B00433).

\ifCLASSOPTIONcaptionsoff
  \newpage
\fi


%
    \bibliographystyle{IEEEtran}
    \bibliography{MAIN}

\end{document}